\documentclass[11pt]{article} 

\usepackage[utf8]{inputenc} 
\usepackage{geometry} 
\usepackage{graphicx} 
\usepackage{booktabs} 
\usepackage{array} 
\usepackage{paralist} 
\usepackage{verbatim} 
\usepackage{subfig} 
\usepackage{fancyhdr} 
\usepackage{sectsty}
\usepackage[nottoc,notlof,notlot]{tocbibind} 
\usepackage[titles,subfigure]{tocloft} 

\usepackage{amsmath}
\usepackage{amssymb}
\usepackage{amsthm}
\usepackage[usenames, dvipsnames]{color}

\allsectionsfont{\sffamily\mdseries\upshape} 

\title{What is meant by~`$P(R\mspace{3mu}|\,Y_{obs})$'?}
\author{JC Galati$^1$}
\date{%
    \today \\
    \mbox{}\\%
    $^1$\small{Department of Mathematics and Statistics, La Trobe University, Melbourne, VIC 3083}
}

\begin{document}
\maketitle

\abstract{
   Missing at Random (MAR) is a central concept in incomplete data methods, and often it is stated as
   $P(R\mspace{3mu}|\,Y_{obs}, Y_{mis}) = P(R\mspace{3mu}|\,Y_{obs})$.
   This notation has been used in the literature for more than three decades and has become the de\,facto standard.
   In some cases, the notation has been misinterpreted to be a statement about conditional independence.
   While previous work has sought to clarify the required definitions, a clear explanation of the standard notation
   seems to be lacking, and a definition of the function $P(R\mspace{3mu}|\,Y_{obs})$ for non-MAR mechanisms is
   difficult to locate in the literature.
   The aim of this paper is to fill these gaps.

   \vspace*{3mm}
}

\section{Introduction}
   \label{Sect:Introduction}
In a foundational work on statistical modeling of incomplete data, a joint distribution for the data variables,~$Y$, and
missingness indicator variables,~$R$, was considered, and conditions under which~$R$ need not be modelled were
identified~(Rubin, 1976). In this work the concept of Missing at Random (MAR) was introduced. A decade later,
Little and Rubin {(1987\;p.\,90; 2002\;p.\,119)} stated the condition as follows:

\vspace*{2mm}
\noindent
\textit{
\quad ``Observe that if the distribution of the missing-data mechanism does not depend on the missing values
$Y_\text{mis}$, that is, if
\begin{equation}
   \mspace{195mu}
   f(R|Y_\text{obs}, Y_\text{mis}, \psi) =
      f(R|Y_\text{obs}, \psi),
   \notag \mspace{135mu} (5.13)
\end{equation}
...\quad Rubin (1976) defines the missing data to be missing at random (MAR) when Eq. (5.13) is satisfied.''
}

\vspace*{2mm}
\noindent
{Schafer (1997\;p.\,10)}  adopted similar notation: ``$P(R|\,Y_{obs}, Y_{mis}, \xi) = P(R|\,Y_{obs}, \xi)$''.

There has been confusion in the literature around how to interpet equation (5.13) correctly, and Mealli and
Rubin\;(2015) pointed out that MAR is not a statement about conditional independence. One factor that is likely to
have contributed to this confusion is that neither `$P(R|Y_{obs}, Y_{mis})$' or~`$P(R |\, Y_{obs})$' in the standard
equation are to be be interpreted as probability distributions if one is to arrive at a correct statement of the MAR
condition, and this has not been made clear in the literature. Another is that while the definition of `$P(R |\, Y_{obs})$'
is clear for MAR mechanisms, a definition for non-MAR missingness mechanisms is difficult to locate, and in the absense
of such a definition, (5.13) cannot be considered to give a definition for MAR, nor can it be related back to the
definition framed in Rubin\;(1976).

The purpose of this paper is to explain how~$P(R|Y_{obs}, Y_{mis})$ and $P(R|Y_{obs})$ are to be interpreted in
the statement `$P(R|Y_{obs}, Y_{mis}) = P(R|Y_{obs})$', which we do using the analogy of the relationship between
a likelihood function and a model of probability densities, and to give a simple-to-understand {definition} of
`$P(R|Y_{obs})$' for all missingness mechanisms (MAR and non-MAR) so that the standard notation states a
mathematical condition equivalent to the condition defined by Rubin\;(1976).

\section{Notation for $(Y, R)$}
    \label{Sect:Notation}
Sections~\ref{Sect:RandomVectors} to~\ref{Sect:Densities} below summarise notation given in Galati\;(2019) and
are included here for completeness.

\subsection{Random Vectors}
    \label{Sect:RandomVectors}

Throughout, $Y$ denotes a random vector modelling the observed and unobserved data comprising all units in the
study jointly, and $R$ denotes a random vector of binary response random variables of the same dimension as~$Y$,
where `1' means observed. Joint distributions for the pair of random vectors $(Y, R)$ will be referred to as
\textbf{full distributions}.

\vspace*{2mm}
\noindent
\textbf{Note.}
We have no need to treat vectors as denoting column matrices.
\hfill $\qedsymbol$

\vspace*{2mm}
\noindent
\textbf{Note.}
Typically a data analyst thinks of a given $\mathbf{y}$ as comprising a rectangular matrix with each column
pertaining to a specific `variable' (for example, blood pressure) and each row pertaining to a specific unit
(for example, an individual in the study). In our notation, the data matrix is shaped so that there is a single row
with the data for the various units placed side by side in sets of colulmns.
\hfill $\qedsymbol$

\subsection{Sample Spaces}
    \label{Sect:SampleSpaces}
Let $\mathcal{R} = \{\mathbf{r}_1, \mathbf{r}_2, \ldots, \mathbf{r}_k \}$ be the set of distinct missingness
patterns with $\mathbf{r}_1 = \mathbf{1}$ denoting the `all ones' vector corresponding to the complete cases.
For convenience, we let $\mathbf{r}_0 = \mathbf{0}$ denote the `all zeros' vector corresponding to
non-participants, where it may or may not be the case that $\mathbf{r}_j = \mathbf{r}_0$ for some
$j\in\{1, 2, \ldots, k\}$.
(We exclude $j=0$ so as to avoid ever having $P(\mathbf{r}_0) = 0$.)
Note that the dot product $\mathbf{r}_j\boldsymbol{\cdot}\mathbf{r}_j$ gives the number of values observed
when the $j^{th}$ response pattern is realised and, in particular, $\mathbf{r}_1\boldsymbol{\cdot}\mathbf{r}_1$
gives the number of variables in~$R$ (and also in~$Y$).
Let $\mathcal{Y} = \text{range}(Y)$ be the set of realisable datasets, where a \textbf{realizable} dataset contains
complete data including all values that may or may not be observable.

Let
$\Omega = \mathcal{Y}\times\mathcal{R} =
   \Omega_1 \;\dot{\cup}\; \Omega_2 \;\dot{\cup}\; \cdots \;\dot{\cup}\; \Omega_k$
be the \textbf{full sample space} of realisable pairs of datasets and missingness patterns,
where $\Omega_j = \mathcal{Y}\times\{\mathbf{r}_j\}$ for~$\mathbf{r}_j\in\mathcal{R}$.
When the subscript $j$ of $\mathbf{r}$ is omitted, we denote $\Omega_j$ by~$\Omega_\mathbf{r}$.
Let $\pi_Y$ and $\pi_R$ denote the projections $(\mathbf{y}, \mathbf{r})\mapsto\mathbf{y}$ and
$(\mathbf{y}, \mathbf{r})\mapsto\mathbf{r}$, respectively.

\subsection{Projections on $\mathcal{Y}$ and~$\Omega_j$}
    \label{Sect:Projections}
For $j=1, 2, \ldots, k$, let $\pi(\mathbf{r}_j)\,:\,\mathcal{Y}\rightarrow\mathcal{Y}^{\pi(\mathbf{r}_j)}$ and
$\pi(\neg\mathbf{r}_j)\,:\,\mathcal{Y}\rightarrow\mathcal{Y}^{\pi(\neg\mathbf{r}_j)}$
denote the projections extracting from each $\mathbf{y}$ vector the vectors of its observed and unobserved
values, respectively, according to the missingness pattern $\mathbf{r}_j$.
(In logic, `$\neg$' is commonly used for negation.)
By convention we set $\pi(\mathbf{r}_0) = \pi(\neg\mathbf{r}_1) = \varnothing$.

To apply these projections correctly over $\Omega$, we define the following mappings
$o\,:\,\mathcal{R} \rightarrow
      \{ \pi(\mathbf{r})\circ\pi_Y : \Omega_{\mathbf{r}}\rightarrow\mathcal{Y}^{\pi(\mathbf{r})}\}$
and 
$m\,:\,\mathcal{R} \rightarrow
      \{ \pi(\neg\mathbf{r})\circ\pi_Y :
      \Omega_{\mathbf{r}}\rightarrow\mathcal{Y}^{\pi(\neg\mathbf{r})} \}$
and use an abbreviated notation to refer to the images of $(\mathbf{y}, \mathbf{r})\in\Omega$ under these
mappings:
\begin{alignat}{1}
   \mathbf{y}^{ob(\mathbf{r})} \,&:=\,
      (\mathbf{y}, \mathbf{r})^{o(\pi_R(\mathbf{y}, \mathbf{r}))}
   \label{Eq:YObR} \\
   \mathbf{y}^{mi(\mathbf{r})} \,&:=\,
      (\mathbf{y}, \mathbf{r})^{m(\pi_R(\mathbf{y}, \mathbf{r}))}.
   \label{Eq:YMiR}
\end{alignat}
Additionally, for $\mathbf{r}\in\mathcal{R}$ and $\mathbf{y}\in\mathcal{Y}$ set
\begin{equation}
   \mathbf{y}^{ot(\mathbf{r})} \,:=\, 
   \begin{cases} 
      \mathbf{y}^{\pi(\mathbf{r})}
         & \text{over } \mathcal{Y} \\
      (\mathbf{y}, \mathbf{r}_j)^{o(\pi_R(\mathbf{y}, \mathbf{r}))}
         & \text{over } \mathcal{Y}\times\mathcal{R}
   \end{cases}
   \label{Eq:YObtR}
\end{equation}
and
\begin{equation}
   \mathbf{y}^{mt(\mathbf{r})} \,:=\, 
   \begin{cases} 
      \mathbf{y}^{\pi(\neg\mathbf{r})}
         & \text{over } \mathcal{Y} \\
      (\mathbf{y}, \mathbf{r}_j)^{m(\pi_R(\mathbf{y}, \mathbf{r}))}
         & \text{over } \mathcal{Y}\times\mathcal{R}.
   \end{cases}
   \label{Eq:YMitR}
\end{equation}

\vspace*{2mm}
\noindent
\textbf{Note.}
In the notations in (\ref{Eq:YObR})$-$(\ref{Eq:YMitR}) only the four symbols $\mathbf{y}^{ob(\mathbf{r})}$,
$\mathbf{y}^{mi(\mathbf{r})}$, $\mathbf{y}^{ot(\mathbf{r})}$ and~$\mathbf{y}^{mt(\mathbf{r})}$ are
required for working with densities for the distributions for~$(Y, R)$ themselves.
\hfill $\qedsymbol$

\vspace*{2mm}
\noindent
\textbf{Note.}
The vectors $\mathbf{y}^{ob(\mathbf{r})}$ and $\mathbf{y}^{ot(\mathbf{r})}$ have length
$\mathbf{r}\boldsymbol{\cdot}\mathbf{r}$ while the vectors $\mathbf{y}^{mi(\mathbf{r})}$ and
$\mathbf{y}^{mt(\mathbf{r})}$ have length
$\mathbf{r}_1\boldsymbol{\cdot}\mathbf{r}_1 - \mathbf{r}\boldsymbol{\cdot}\mathbf{r}$.
Note that these lengths vary from missingness pattern to missingness pattern.
\hfill $\qedsymbol$

\subsection{Observable Data Events}
    \label{Sect:ODEs}
Given $(\mathbf{y}, \mathbf{r})\in\mathcal{Y}\times\mathcal{R}$, we call
\begin{equation}
   \Omega_{(\mathbf{y},\mathbf{r})} \;=\;
      \{\, (\mathbf{y}_*, \mathbf{r}) \,:\, \mathbf{y}_*^{ob(\mathbf{r})}=\mathbf{y}^{ob(\mathbf{r})}\,\}
      \;\subseteq\; \mathcal{Y}\times\mathcal{R}
   \label{Eq:ODE}
\end{equation}
the \textbf{observed data event} corresponding to~$(\mathbf{y}, \mathbf{r})$.
The set $\Omega_{(\mathbf{y},\mathbf{r})}$ consists of all pairs
$(\mathbf{y}_*,\mathbf{r})$ where $\mathbf{y}_*$ is a complete dataset having the same observed values
as~$\mathbf{y}$ (as defined by the response pattern~$\mathbf{r}$). For a fixed~$\mathbf{r}\in\mathcal{R}$,
the events in~(\ref{Eq:ODE}) partition~$\Omega_\mathbf{r}$, and over all~$\mathbf{r}$ they give a partition
of~$\Omega$. These \textbf{observable data events} are the classes of the equivalence relation defined by setting
for all $(\mathbf{y}_1, \mathbf{r}_1), (\mathbf{y}_2, \mathbf{r}_2) \in \mathcal{Y}\times\mathcal{R}$,
$(\mathbf{y}_1, \mathbf{r}_1) \sim_\text{ob} (\mathbf{y}_2, \mathbf{r}_2)$ if, and only if,
$\mathbf{r}_1 = \mathbf{r}_2$ and $\mathbf{y}_1^{ob(\mathbf{r}_1)} = \mathbf{y}_2^{ob(\mathbf{r}_2)}$.

\subsection{Density Functions}
    \label{Sect:Densities}
We specify full distributions for $(Y, R)$ through density functions $h:\Omega\rightarrow\mathbb{R}$,
with probabilities determined by integration: $P(A) = \int_A h$ for any $A\subseteq\mathcal{Y}\times\mathcal{R}$
for which a probability can be defined (see Ash and Dol\'{e}ans-Dade\;(2000) or~Shorack\;(2000) for details). Note
{that} we suppress the dominating measure in the notation. Two different ways of factorizing $h$ are useful:
\begin{equation}
   h(\mathbf{y}, \mathbf{r}) \; = \;
      f(\mathbf{y})\,g(\mathbf{r}\,|\,\mathbf{y}) \; = \;
      p(\mathbf{r})\,p(\mathbf{y}\,|\,\mathbf{r})
   \label{Eq:FullDensity}
\end{equation}
for all $(\mathbf{y},\mathbf{r})\in\mathcal{Y}\times\mathcal{R}$. The first factorization is called a \textbf{selection
model} factorization of~$h$, and the factor $g(\mathbf{r}\,|\,\mathbf{y})$ is called a \textbf{missingness mechanism}.
The second factorization is called a \textbf{pattern-mixture} factorization, and for each~$\mathbf{r}\in\mathcal{R}$,
we call the conditional density $p(\mathbf{y}\,|\,\mathbf{r})$ the \textbf{pattern mixture component} pertaining
to~$\mathbf{r}$.

\vspace*{2mm}
\noindent
\textbf{Note.}
As specified in~(\ref{Eq:FullDensity}), a missingness mechanism $g(\mathbf{r}\,|\,\mathbf{y})$ is a function of two
vector variables $\mathbf{y}$ and $\mathbf{r}$ defined on all of $\mathcal{Y}\times\mathcal{R}$ subject to the
restrictions that
$0\leq g(\mathbf{r}\,|\,\mathbf{y})\leq 1$ for each $(\mathbf{y},\mathbf{r})\in\mathcal{Y}\times\mathcal{R}$
and $\sum_{i=1}^{k}g(\mathbf{r}_i\,|\,\mathbf{y})=1$ for each fixed $\mathbf{y}\in\mathcal{Y}$.
We stress that instead of the ususal interpretation of considering a missingness mechanism to give a conditional
probability distribution for~$R$ for each fixed $\mathbf{y}\in\mathcal{Y}$, the perspective that will be relevant for
us is to consider the behaviour of $g$ as a mathematical function of both $\mathbf{y}$ and $\mathbf{r}$ when its
domain, $\mathcal{Y}\times\mathcal{R}$, is restricted to an observed data event
$\Omega_{(\mathbf{y},\mathbf{r})}\subset\mathcal{Y}\times\mathcal{R}$,
that is, to a set of the form~(\ref{Eq:ODE}). This perspective is specific to the incomplete data setting, and typically
does not arise when considering complete-data statistical methods.
\hfill $\qedsymbol$

\vspace*{2mm}
\noindent
\textbf{Note.}
Technically, the symbols $h$, $f$, $g$ and $p$ denote functions and $h(\mathbf{y}, \mathbf{r})$,
$f(\mathbf{y})$, $g(\mathbf{r}\,|\,\mathbf{y})$, $p(\mathbf{r})$ and $p(\mathbf{y}\,|\,\mathbf{r})$ denote real
numbers.
Because it is common in statistics to use the same symbol to denote different densities, for example a joint
density $f(x_1, x_2)$ and a marginal density~$f(x_1)$, we adopt the usual convention and often refer to these
functions by their values.
\hfill $\qedsymbol$

\subsection{Graphical illustration of a full density}
    \label{Sect:Graphical}
Figure~1 provides a pictorial description of a full density and its selection-model and pattern-mixture factorizations.
For graphical simplicity the distributions of the $\mathbf{y}$ vectors are depicted as one-dimensional, but in practice
these distributions are multi-dimensional.

The marginal probabilities $p(\mathbf{r}_j)$ for the $k$ missingness patterns give the marginal distribution for~$R$.
These must sum to~1. The marginal density $f(\mathbf{y})$ is the average of the patten-mixture components
$p(\mathbf{y}\,|\,\mathbf{r}_j)$ for $j=1,\ldots,k$ weighted by their marginal probabilities~$p(\mathbf{r}_j)$.
The missingness mechanism evaluated at a fixed $\mathbf{y}$ vector gives the probability distribution for the $k$
response patterns corresponding to that particular $\mathbf{y}$ vector. These probabilities may vary as $\mathbf{y}$
varies, but for any fixed~$\mathbf{y}$ vector they always sum to~1. The histogram below $f(\mathbf{y})$ depicts
the marginal distribution for $Y$ that would be observable if the data values could always be observed (that is, if missing
data were not possible).

The rectangular regions labelled $\Omega_j$ depict the stratification of $\mathcal{Y}\times\mathcal{R}$ by missingness
pattern. The histogram below each pattern-mixture component $p(\mathbf{y}\,|\,\mathbf{r}_j)$ illustrates the
distribution of the $\mathbf{y}$ vectors in each stratum. The differing distributional shapes across the pattern-mixture
components illustrates the distributional effect of missingness on the complete data (observable and unobservable data
values) \textbf{before} any loss of information is incurred due to some data values being unobservable. For example,
the shape of the density $p(\mathbf{y}\,|\,\mathbf{r}_1)$ depicts the distribution of the complete cases, which in
general will differ from the shape of the marginal density~$f(\mathbf{y})$. This differing shape explains the potential
bias that can result by restricting analyses to complete cases only.

\vspace*{3mm}
\begin{center}
\begin{picture}(350, 270)
   \put(0, 12){
      \thinlines
      \put(250, 15){
         \thinlines
         \put(29, 22){\textbf{The Marginal}}
         \put(30, 11){\textbf{Distributions}}
         \put(33, 0){\textbf{for $R$ and~$Y$}}
         \put(12, 8){$\boldsymbol{\leftarrow}$}
         \put(68, 43){$\boldsymbol{\uparrow}$}
      }
      \put(10, 0){
         \put(0, 12){
            \put(0, 25){$\mathcal{R}$}
            \thinlines
            \multiput(0, 0)(0, 21){2}{\line(1, 0){230}}
            \multiput(0, 0)(230, 0){2}{\line(0, 1){21}}
            \put(23, 4){
               \put(0, 0){        
                  \put(0, 0){$\mathbf{r}_1$}
                  \put(60, 0){$\mathbf{r}_2$}
                  \put(180, 0){$\mathbf{r}_k$}
               }
               \put(2, 10){      
                  \put(0, 0){\circle*{3}}
                  \put(60, 0){\circle*{3}}
                  \put(180, 0){\circle*{3}}
               }
            }
         }
         \put(16, 37){        
            \put(0, 0){$p(\mathbf{r}_1)$}
            \put(35, 0){$+$}
            \put(60, 0){$p(\mathbf{r}_2)$}
            \put(95, 0){$+$}
            \put(122, 0){$\cdots$}
            \put(155, 0){$+$}
            \put(180, 0){c($\mathbf{r}_k)$}
            \put(211, 0){$= 1$}
         }
      }
      \put(295, 65){
         \put(0, 19){
            \put(0, 84){$\mathcal{Y}$}
            \thinlines
            \multiput(0, 11)(0, 70){2}{\line(1, 0){50}}
            \multiput(0, 11)(50, 0){2}{\line(0, 1){70}}
            \thinlines
            \multiput(12, 11)(0, 2){35}{\put(0, 0){\line(0, 1){1.3}}}
            \put(9, 0){$\mathbf{y}$}
         }
         \put(15, 152){$f(\mathbf{y})$}
         \put(0, 117){
            \thinlines
            \put(0, 0){\line(1, 0){50}}
            \put(0, 5){\line(1, 0){5}}
            \put(0, 0){\multiput(0, 0)(5, 0){2}{\line(0, 1){5}}}
            \put(5, 12){\line(1, 0){5}}
            \put(5, 0){\multiput(0, 0)(5, 0){2}{\line(0, 1){12}}}
            \put(10, 18){\line(1, 0){5}}
            \put(10, 0){\multiput(0, 0)(5, 0){2}{\line(0, 1){18}}}
            \put(15, 22){\line(1, 0){5}}
            \put(15, 0){\multiput(0, 0)(5, 0){2}{\line(0, 1){22}}}
            \put(20, 27){\line(1, 0){5}}
            \put(20, 0){\multiput(0, 0)(5, 0){2}{\line(0, 1){27}}}
            \put(25, 25){\line(1, 0){5}}
            \put(25, 0){\multiput(0, 0)(5, 0){2}{\line(0, 1){25}}}
            \put(30, 19){\line(1, 0){5}}
            \put(30, 0){\multiput(0, 0)(5, 0){2}{\line(0, 1){19}}}
            \put(35, 15){\line(1, 0){5}}
            \put(35, 0){\multiput(0, 0)(5, 0){2}{\line(0, 1){15}}}
            \put(40, 11){\line(1, 0){5}}
            \put(40, 0){\multiput(0, 0)(5, 0){2}{\line(0, 1){11}}}
            \put(45, 7){\line(1, 0){5}}
            \put(45, 0){\multiput(0, 0)(5, 0){2}{\line(0, 1){7}}}
         }
      }
      \put(269, 125){
         \put(2.5, 8){$\pi_\mathcal{Y}$}
         \put(0, 0){$\boldsymbol{\longmapsto}$}
      }
      \put(0, 55){
         \thinlines
         \multiput(0, 0)(0, 204){2}{\multiput(0, 0)(2, 0){130}{\line(1, 0){1.2}}}
         \multiput(0, 0)(260, 0){2}{\multiput(0, 0)(0, 2){102}{\line(0, 1){1.2}}}
      }
      \put(10, 65){
         \put(55, 175){\large{\textbf{The Full Distribution}}\normalsize}
         \put(0, 30){
            \put(0, 73){
               \put(0, 0){$\Omega_1$}
               \put(60, 0){$\Omega_2$}
               \put(180, 0){$\Omega_k$}
            }
            \thinlines
            \multiput(0, 0)(60, 0){2}{
               \multiput(0, 0)(0, 70){2}{\line(1, 0){50}}
               \multiput(0, 0)(50, 0){2}{\line(0, 1){70}}
            }
            \put(140, 35){$\cdots$}
            \put(180, 0){
               \multiput(0, 0)(0, 70){2}{\line(1, 0){50}}
               \multiput(0, 0)(50, 0){2}{\line(0, 1){70}}
            }
            \put(12, 0){
               \linethickness{0.5mm}
               \put(0, 0){\line(0, 1){70}}
               \put(60, 0){\line(0, 1){50}}
               \put(180, 0){\line(0, 1){30}}
               \thinlines
            }
            \put(16, 0){
               \put(0, 32){$\mathbf{y}^{ob(\mathbf{r}_1)}$}
               \put(60, 56){$\mathbf{y}^{mi(\mathbf{r}_2)}$}
               \put(60, 22){$\mathbf{y}^{ob(\mathbf{r}_2)}$}
               \put(180, 45){$\mathbf{y}^{mi(\mathbf{r}_k)}$}
               \put(180, 12){$\mathbf{y}^{ob(\mathbf{r}_k)}$}
            }
         }
         \put(1, 19){
            \put(0, 0){$(\mathbf{y}, \mathbf{r}_1)$}
            \put(60, 0){$(\mathbf{y}, \mathbf{r}_2)$}
            \put(180, 0){$(\mathbf{y}, \mathbf{r}_k)$}
         }
         \put(8, 0){
            \put(0, 0){$g(\mathbf{r}_1\,|\,\mathbf{y})$}
            \put(43, 0){$+$}
            \put(60, 0){$g(\mathbf{r}_2\,|\,\mathbf{y})$}
            \put(105, 0){$+$}
            \put(132, 0){$\cdots$}
            \put(165, 0){$+$}
            \put(180, 0){$g(\mathbf{r}_k\,|\,\mathbf{y})$}
            \put(219, 0){$=1$}
         }
         \put(8, 150){
            \put(0, 0){$p(\mathbf{y}\,|\,\mathbf{r}_1)$}
            \put(60, 0){$p(\mathbf{y}\,|\,\mathbf{r}_2)$}
            \put(180, 0){$p(\mathbf{y}\,|\,\mathbf{r}_k)$}
         }
         \put(0, 117){
            \thinlines
            \put(0, 0){
               \put(0, 0){\line(1, 0){50}}
               \put(0, 2){\line(1, 0){5}}
               \put(0, 0){\multiput(0, 0)(5, 0){2}{\line(0, 1){2}}}
               \put(5, 5){\line(1, 0){5}}
               \put(5, 0){\multiput(0, 0)(5, 0){2}{\line(0, 1){5}}}
               \put(10, 10){\line(1, 0){5}}
               \put(10, 0){\multiput(0, 0)(5, 0){2}{\line(0, 1){10}}}
               \put(15, 20){\line(1, 0){5}}
               \put(15, 0){\multiput(0, 0)(5, 0){2}{\line(0, 1){20}}}
               \put(20, 23){\line(1, 0){5}}
               \put(20, 0){\multiput(0, 0)(5, 0){2}{\line(0, 1){23}}}
               \put(25, 22){\line(1, 0){5}}
               \put(25, 0){\multiput(0, 0)(5, 0){2}{\line(0, 1){22}}}
               \put(30, 17){\line(1, 0){5}}
               \put(30, 0){\multiput(0, 0)(5, 0){2}{\line(0, 1){17}}}
               \put(35, 15){\line(1, 0){5}}
               \put(35, 0){\multiput(0, 0)(5, 0){2}{\line(0, 1){15}}}
               \put(40, 12){\line(1, 0){5}}
               \put(40, 0){\multiput(0, 0)(5, 0){2}{\line(0, 1){12}}}
               \put(45, 7){\line(1, 0){5}}
               \put(45, 0){\multiput(0, 0)(5, 0){2}{\line(0, 1){7}}}
            }
            \put(60, 0){
               \put(0, 0){\line(1, 0){50}}
               \put(0, 12){\line(1, 0){5}}
               \put(0, 0){\multiput(0, 0)(5, 0){2}{\line(0, 1){12}}}
               \put(5, 18){\line(1, 0){5}}
               \put(5, 0){\multiput(0, 0)(5, 0){2}{\line(0, 1){18}}}
               \put(10, 10){\line(1, 0){5}}
               \put(10, 0){\multiput(0, 0)(5, 0){2}{\line(0, 1){10}}}
               \put(15, 20){\line(1, 0){5}}
               \put(15, 0){\multiput(0, 0)(5, 0){2}{\line(0, 1){20}}}
               \put(20, 23){\line(1, 0){5}}
               \put(20, 0){\multiput(0, 0)(5, 0){2}{\line(0, 1){23}}}
               \put(25, 22){\line(1, 0){5}}
               \put(25, 0){\multiput(0, 0)(5, 0){2}{\line(0, 1){22}}}
               \put(30, 15){\line(1, 0){5}}
               \put(30, 0){\multiput(0, 0)(5, 0){2}{\line(0, 1){15}}}
               \put(35, 10){\line(1, 0){5}}
               \put(35, 0){\multiput(0, 0)(5, 0){2}{\line(0, 1){10}}}
               \put(40, 6){\line(1, 0){5}}
               \put(40, 0){\multiput(0, 0)(5, 0){2}{\line(0, 1){6}}}
               \put(45, 2){\line(1, 0){5}}
               \put(45, 0){\multiput(0, 0)(5, 0){2}{\line(0, 1){2}}}
            }
            \put(180, 0){
               \put(0, 0){\line(1, 0){50}}
               \put(0, 2){\line(1, 0){5}}
               \put(0, 0){\multiput(0, 0)(5, 0){2}{\line(0, 1){2}}}
               \put(5, 3){\line(1, 0){5}}
               \put(5, 0){\multiput(0, 0)(5, 0){2}{\line(0, 1){3}}}
               \put(10, 5){\line(1, 0){5}}
               \put(10, 0){\multiput(0, 0)(5, 0){2}{\line(0, 1){5}}}
               \put(15, 18){\line(1, 0){5}}
               \put(15, 0){\multiput(0, 0)(5, 0){2}{\line(0, 1){18}}}
               \put(20, 14){\line(1, 0){5}}
               \put(20, 0){\multiput(0, 0)(5, 0){2}{\line(0, 1){14}}}
               \put(25, 15){\line(1, 0){5}}
               \put(25, 0){\multiput(0, 0)(5, 0){2}{\line(0, 1){15}}}
               \put(30, 23){\line(1, 0){5}}
               \put(30, 0){\multiput(0, 0)(5, 0){2}{\line(0, 1){23}}}
               \put(35, 17){\line(1, 0){5}}
               \put(35, 0){\multiput(0, 0)(5, 0){2}{\line(0, 1){17}}}
               \put(40, 11){\line(1, 0){5}}
               \put(40, 0){\multiput(0, 0)(5, 0){2}{\line(0, 1){11}}}
               \put(45, 6){\line(1, 0){5}}
               \put(45, 0){\multiput(0, 0)(5, 0){2}{\line(0, 1){6}}}
            }
         }
      }
      \thinlines
   }
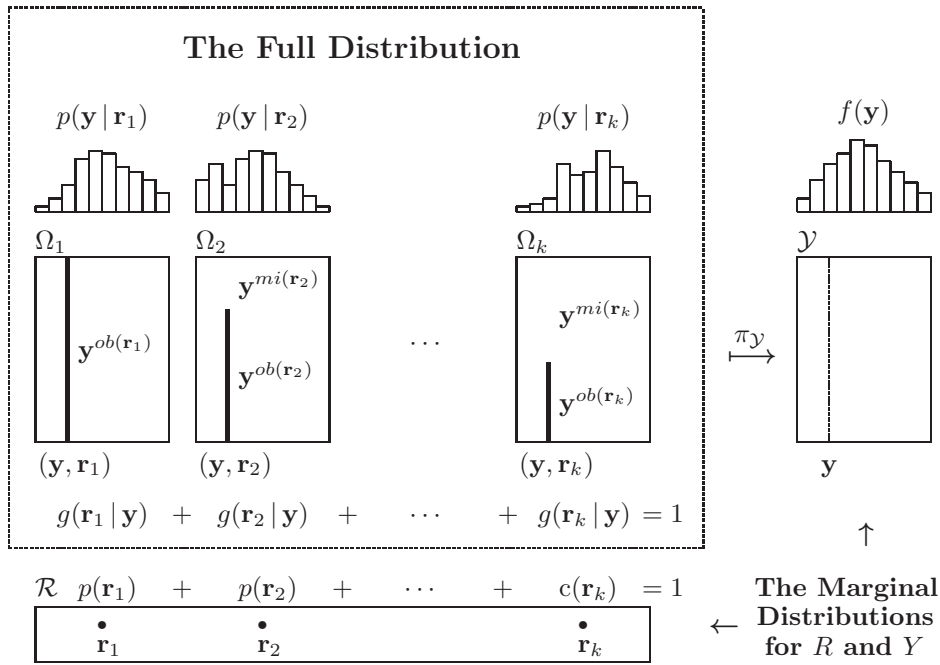
   \put(0, 2){\textbf{Figure~1.} Selection-model and pattern-mixture factorisations of a full density.}
\end{picture}
\end{center}

\vspace*{2mm}
The vertical bar in each of the sets $\Omega_j$ and in $\mathcal{Y}$ represents a single dataset
($\mathbf{y}$~vector). In the sets $\Omega_j$ the bar is partitioned into a thick black part representing the values
of $\mathbf{y}$ that are always observed, and a white unseen part representing the values of $\mathbf{y}$ that are
never observed (whenever the missingness pattern $\mathbf{r}_j$ is realised). The dotted vertical line in
$\mathcal{Y}$ represents the fact that $\mathbf{y}$ values in the marginal distribution for $Y$ cannot be separated
into observed and missing parts, and represent a mixture of observed and missing values averaged over all
missingness patterns. In the figure, and in later notation, we assume that we can reorder the entries of $\mathbf{y}$
as is convenient to separate the values into missing and observed parts, and that we know how to reverse this
reordering to compare $\mathbf{y}$ vectors across the sets $\Omega_j$ and~$\mathcal{Y}$.

\section{Missingness at Random}
    \label{Sect:MAR}
Missing at Random (MAR) is a property of a missingness mechanism postulated to hold when the domain of the
missingness mechanism, $\mathcal{Y}\times\mathcal{R}$, is restricted to a specific observable data
event,~$\Omega_{(\mathbf{y},\mathbf{r})}$.
One of these events is illustrated in Figure~2. The observable data event consists of all datasets lying between the
dotted vertical lines. Note that~$\mathbf{y}_*^{ob(\mathbf{r})} = \mathbf{y}^{ob(\mathbf{r})} $ is the same for
all datasets~$\mathbf{y}_*$ in the event, but each~$\mathbf{y}_*^{mi(\mathbf{r})}$ vector is different. The
definition of MAR was framed by Rubin\;(1976). Here we state it in a slightly different form.

\vspace*{2mm}
\noindent
\textbf{Definition~\ref{Sect:MAR}.1.}
Given $h(\mathbf{y}, \mathbf{r}) = f(\mathbf{y})\,g(\mathbf{r}\,|\,\mathbf{y})$ factorised in selection model form
together with observed data~$\Omega_{(\mathbf{y},\mathbf{r})}$,
we say that the missingness mechanism $g$ is \textbf{missing at random} (\textbf{MAR}) with respect to 
$\Omega_{(\mathbf{y},\mathbf{r})}$ if $g$ is constant
on~$\Omega_{(\mathbf{y},\mathbf{r})}$.
\hfill $\qedsymbol$

\vspace*{3mm}
\begin{center}
\begin{picture}(210, 118)
   \put(0, 15){
      \put(30, 85){$\Omega_\mathbf{r}$}
      \thinlines
      \multiput(0, 47)(198, 0){2}{$\cdots$}
      \put(30, 12){
         \multiput(0, 0)(0, 70){2}{\line(1, 0){150}}
         \multiput(0, 0)(150, 0){2}{\line(0, 1){70}}
      }
      \linethickness{0.5mm}
      \put(76, 12){\line(0, 1){50}}
      \thinlines
      \multiput(60, 12)(50, 0){2}{\multiput(0, 0)(0, 2){35}{\line(0, 1){1.2}}}
      \put(80, 68){$\mathbf{y}_*^{mi(\mathbf{r})}$}
      \put(80, 35){$\mathbf{y}^{ob(\mathbf{r})}$}
      \put(70, 2){$(\mathbf{y}, \mathbf{r})$}
      \put(73, 92){$\Omega_{(\mathbf{y},\mathbf{r})}$}
      \put(60, 92){$\leftarrow$}
      \put(101, 92){$\rightarrow$}
      \multiput(60, 86)(50, 0){2}{\multiput(0, 0)(0, 2){8}{\line(0, 1){1.2}}}
      \thinlines
   }
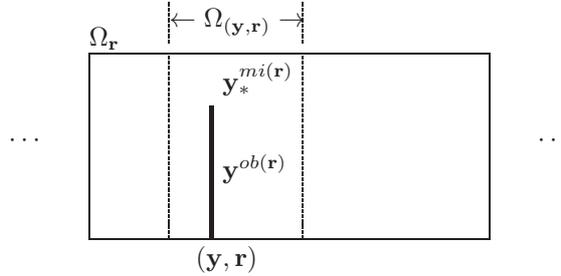
   \put(0, 2){\textbf{Figure~2.} An observable data event in $\Omega_\mathbf{r}\subset\mathcal{Y}\times\mathcal{R}$.}
\end{picture}
\end{center}

\vspace*{2mm}

\vspace*{2mm}
\noindent
\textbf{{Note}.}
MAR as we have defined it is equivalent to the definition framed by Rubin\;(1976), except that we have identified the
property as an attribute of the missingness mechanism of a single full density. Rubin framed the definition for a model
of densities $\mathcal{M} = \{ h_{(\theta, \psi)} \,:\, (\theta, \psi)\in\Delta\subseteq\Theta\times\Psi\,\}$, and this can
be accommodated by requiring that MAR hold with respect to
$\Omega_{(\mathbf{y},\mathbf{r})}$ for all densities
$h_{(\theta, \psi)}$ in~$\mathcal{M}$. Everywhere MAR (Seaman\;et.\,al.,\,2013) is accommodated simply by
requiring that MAR hold with respect to all observable data events (for all densities in~$\mathcal{M}$).
\hfill $\qedsymbol$

\vspace*{2mm}
\noindent
\textbf{{Note}.}
The terminology chosen by Rubin (1976) attributes MAR to the data: \textit{``The data ... are missing at random ...''}.
We have attributed it to the missingness mechanism because it is possible to have densities $h_1$ and~$h_2$
with~$h_1$ MAR and~$h_2$ not MAR with respect to the same
event~$\Omega_{(\mathbf{y},\mathbf{r})}\subset\mathcal{Y}\times\mathcal{R}$, so the property is not an attribute
of the realised data.  Nevertheless, Rubin's terminology makes sense from the following perspective. It can be shown
that when MAR holds with respect to~$\Omega_{(\mathbf{y},\mathbf{r})}$, then
$p(\mathbf{y}_{\hspace{-0.3mm}*}^{mi(\mathbf{r})} |\, \mathbf{y}^{ob(\mathbf{r})}, \mathbf{r}) =
   f(\mathbf{y}_{\hspace{-0.3mm}*}^{mi(\mathbf{r})} |\, \mathbf{y}^{ob(\mathbf{r})})$
for all
$(\mathbf{y}^{ob(\mathbf{r})}, \mathbf{y}_{\hspace{-0.3mm}*}^{mi(\mathbf{r})}, \mathbf{r}) \in
   \Omega_{(\mathbf{y},\mathbf{r})}$.
This equality says that under MAR, datasets $\mathbf{y}_*$ with a fixed pattern of missingness~$\mathbf{r}$ and
a fixed set of observed values~$\mathbf{y}_*^{ob(\mathbf{r})} = \mathbf{y}^{ob(\mathbf{r})}$ can be drawn
`at random' from~$\Omega_{(\mathbf{y},\mathbf{r})}$.
(see Galati (2019,\,Appendix\;D), for example).
 \hfill $\qedsymbol$

\vspace*{2mm}
\noindent
\textbf{{Note}.}
A pictorial way to interpret the effect of missingness on the distribution of the $\mathbf{y}$ values is illustrated in
Figure~3.

\vspace*{3mm}
\begin{center}
\begin{picture}(350, 178)
   \put(0, 15){
      \put(0, 85){$\Omega_\mathbf{r}$}
      \put(200, 85){$\mathcal{Y}$}
      \thinlines
      \multiput(0, 12)(200, 0){2}{
         \multiput(0, 0)(0, 70){2}{\line(1, 0){150}}
         \multiput(0, 0)(150, 0){2}{\line(0, 1){70}}
      }
      \put(30, 0){
         \linethickness{0.5mm}
         \put(16, 12){\line(0, 1){50}}
     }
      \put(230, 0){
         \linethickness{0.3mm}
         \multiput(16, 12)(0, 2){35}{\line(0, 1){1.2}}
     }
      \multiput(30, 0)(200, 0){2}{
         \thinlines
         \multiput(0, 12)(66, 0){2}{\multiput(0, 0)(0, 2){35}{\line(0, 1){1.2}}}
         \multiput(0, 86)(66, 0){2}{\multiput(0, 0)(0, 2){8}{\line(0, 1){1.2}}}
      }
      \put(52, 68){$\mathbf{y}_*^{mi(\mathbf{r})}$}
      \put(52, 35){$\mathbf{y}^{ob(\mathbf{r})}$}
      \put(254, 68){$\mathbf{y}_*^{mt(\mathbf{r})}$}
      \put(254, 35){$\mathbf{y}^{ot(\mathbf{r})}$}
      \put(39.5, 2){$(\mathbf{y}, \mathbf{r})$}
      \put(243, 2){$\mathbf{y}$}
      \put(51, 92){$\Omega_{(\mathbf{y},\mathbf{r})}$}
      \put(240.2, 92){$\pi_\mathcal{Y}\hspace*{-0.7mm}\left(\Omega_{(\mathbf{y},\mathbf{r})}\right)$}
      \put(30, 92){$\longleftarrow$}
      \put(80.5, 92){$\longrightarrow$}
      \put(230, 92){$\leftarrow$}
      \put(287, 92){$\rightarrow$}
      \put(20, 148){$p(\mathbf{y} |\, \mathbf{r}) \propto f(\mathbf{y})\,g(\mathbf{r}|\,\mathbf{y})$}
      \put(30, 110){
         \thinlines
         \put(0, 0){\line(1, 0){66}}
         \put(0, 2){\line(1, 0){6}}
         \multiput(0, 0)(6, 0){2}{\line(0, 1){2}}
         \put(6, 3){\line(1, 0){6}}
         \multiput(6, 0)(6, 0){2}{\line(0, 1){3}}
         \put(12, 7){\line(1, 0){6}}
         \multiput(12, 0)(6, 0){2}{\line(0, 1){7}}
         \put(18, 12){\line(1, 0){6}}
         \multiput(18, 0)(6, 0){2}{\line(0, 1){12}}
         \put(24, 25){\line(1, 0){6}}
         \multiput(24, 0)(6, 0){2}{\line(0, 1){25}}
         \put(30, 18){\line(1, 0){6}}
         \multiput(30, 0)(6, 0){2}{\line(0, 1){18}}
         \put(36, 14){\line(1, 0){6}}
         \multiput(36, 0)(6, 0){2}{\line(0, 1){14}}
         \put(42, 11){\line(1, 0){6}}
         \multiput(42, 0)(6, 0){2}{\line(0, 1){11}}
         \put(48, 10){\line(1, 0){6}}
         \multiput(48, 0)(6, 0){2}{\line(0, 1){10}}
         \put(54, 7){\line(1, 0){6}}
         \multiput(54, 0)(6, 0){2}{\line(0, 1){7}}
         \put(60, 5){\line(1, 0){6}}
         \multiput(60, 0)(6, 0){2}{\line(0, 1){5}}
      }
      \put(230, 110){
         \put(21, 38){$f(\mathbf{y})$}
         \thinlines
         \put(0, 0){\line(1, 0){66}}
         \put(0, 2.5){\line(1, 0){6}}
         \multiput(0, 0)(6, 0){2}{\line(0, 1){2.5}}
         \put(6, 3.75){\line(1, 0){6}}
         \multiput(6, 0)(6, 0){2}{\line(0, 1){3.75}}
         \put(12, 8.75){\line(1, 0){6}}
         \multiput(12, 0)(6, 0){2}{\line(0, 1){8.75}}
         \put(18, 15){\line(1, 0){6}}
         \multiput(18, 0)(6, 0){2}{\line(0, 1){15}}
         \put(24, 31.25){\line(1, 0){6}}
         \multiput(24, 0)(6, 0){2}{\line(0, 1){31.25}}
         \put(30, 22.5){\line(1, 0){6}}
         \multiput(30, 0)(6, 0){2}{\line(0, 1){22.5}}
         \put(36, 17.5){\line(1, 0){6}}
         \multiput(36, 0)(6, 0){2}{\line(0, 1){17.5}}
         \put(42, 13.75){\line(1, 0){6}}
         \multiput(42, 0)(6, 0){2}{\line(0, 1){13.75}}
         \put(48, 12.5){\line(1, 0){6}}
         \multiput(48, 0)(6, 0){2}{\line(0, 1){12.5}}
         \put(54, 8.75){\line(1, 0){6}}
         \multiput(54, 0)(6, 0){2}{\line(0, 1){8.75}}
         \put(60, 6.25){\line(1, 0){6}}
         \multiput(60, 0)(6, 0){2}{\line(0, 1){6.25}}
      }
      \put(168.5, 51){$\pi_\mathcal{Y}$}
      \put(166.5, 43){$\longmapsto$}
      \put(162, 122.5){Same}
      \put(161.5, 113){shape}
      \put(132, 117){$\longleftarrow$}
      \put(200, 117){$\longrightarrow$}
      \thinlines
   }
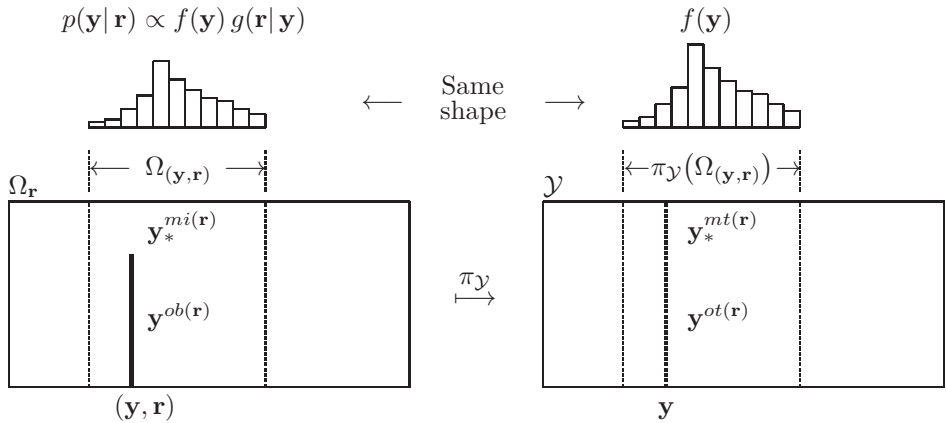
   \put(0, 2){\textbf{Figure~3.} Effect of MAR on pattern-mixture component density.}
\end{picture}
\end{center}

\vspace*{2mm}
\noindent
For a fixed missingness pattern~$\mathbf{r}$, the missingness mechanism $g(\mathbf{r} |\, \mathbf{y})$ considered
as a function of $\mathbf{y}$ alone distorts the shape of the marginal density $f(\mathbf{y})$ to produce the
pattern-mixture component $p(\mathbf{y} |\,\mathbf{r})$ (appropriately scaled). The MAR condition requires that
when restricted to the observable data event $\Omega_{(\mathbf{y},\mathbf{r})}$, the pattern-mixture component
and marginal densities have the same shape (but scaled differently).
\hfill $\qedsymbol$

\vspace*{2mm}
\noindent
\textbf{{Note}.}
The main subtlety of the MAR definition surfaces when one attempts to conceive of a missingness mechansim in
which the probability of a fixed missingness pattern $\mathbf{r}_j$ given by $g(\mathbf{r}_j\,|\,\mathbf{y})$ varies
with the observed values~$\mathbf{y}^{ob(\mathbf{r}_j)}$. Referring back to Figure~1, the equation
$g(\mathbf{r}_1\,|\,\mathbf{y}) + g(\mathbf{r}_2\,|\,\mathbf{y}) + \cdots + g(\mathbf{r}_k\,|\,\mathbf{y}) = 1$
must hold at all times.
So $g(\mathbf{r}_j\,|\,\mathbf{y})$ cannot vary in isolation of the probabilities of the other missingness patterns.
There must be one or more other patterns, say $\mathbf{r}_{i_1},\ldots\mathbf{r}_{i_s}$, whose probabilities in
total vary in the opposite direction to accommodate the change in $g(\mathbf{r}_j\,|\,\mathbf{y})$ for a change
in $\mathbf{y}^{ob(\mathbf{r}_j)}$. If the missingness mechanism is to be everywhere MAR, then MAR must
hold for these offsetting patterns as well, so the probability of the given missingness pattern $\mathbf{r}_j$ can
depend only on the components of $\mathbf{y}^{ob(\mathbf{r}_j)}$ that are defined to be observed by all of
$\mathbf{r}_j,\mathbf{r}_{i_1},\ldots,\mathbf{r}_{i_s}$, and not just those defined to observed by~$\mathbf{r}_j$.
\hfill $\qedsymbol$

\section{Why aren't $P(R|\,Y_{obs}, Y_{mis})$ and $P(R|\,Y_{obs})$ to be treated as probability distributions?}
   \label{Sect:Obstacles}
We use an analogy to likelihood theory to explain the reason why $P(R|\,Y_{obs}, Y_{mis})$ and $P(R|\,Y_{obs})$
are not to be treated as probability distributions in the standard notation for MAR.

Consider a model of densities defined on the real numbers $\mathcal{M} = \{\, f_\theta(x):\theta\in\Theta\,\}.$
There are three different perspectives from which one can view~$\mathcal{M}$.
The natural one is to think of $\Theta$ as the indexing set and each $f_\theta(x)$ as giving a probability distribution
on~$\mathbb{R}$.
A second perspective is to consider $\mathbb{R}$ to be the indexing set and each $L_x(\theta) = f_\theta(x)$
as giving a likelihood function on~$\Theta$.
A third perspective is to consider the entire model as a single function of two variables
$\mathcal{M}:\mathbb{R}\times\Theta\rightarrow\mathbb{R}$ sending $(x, \theta)$ to~$f_\theta(x)$.

The same three perspectives apply to the missingness mechanism $g(\mathbf{r}|\,\mathbf{y})$.
As noted in Section~\ref{Sect:Densities}, for each fixed $\mathbf{y}\in\mathcal{Y}$, one can consider the
missingness mechanism to give a function from $\mathcal{R}$ to $[0,1]$ defined by
$\mathbf{r}\mapsto g(\mathbf{r}\,|\,\mathbf{y})$ for all $\mathbf{r}\in\mathcal{R}$ (that is, as giving a
conditional probability distribution for~$R$ for each fixed $\mathbf{y}\in\mathcal{Y}$). Alternatively, for each
fixed $\mathbf{r}\in\mathcal{R}$, one can consider the missingness mechanism to give a function from
$\mathcal{Y}$ to $[0,1]$ defined by $\mathbf{y}\mapsto g(\mathbf{r}\,|\,\mathbf{y})$ for all
$\mathbf{y}\in\mathcal{Y}$. Or one can consider the missingness mechanism to be a single function (of two vector
variables) from $\Omega$ to $[0,1]$ defined by $(\mathbf{y}, \mathbf{r})\mapsto g(\mathbf{r}\,|\,\mathbf{y})$.

It is the second perspective that is used to interpret the standard notation correctly, \textbf{not} the first.
Specifically, the notation `$P(R|Y_{obs},Y_{mis})$' on the left hand side of the standard equation should be
considered to denote a function of $Y$ with $R$ held fixed, and \textbf{not} a conditional probability distribution
for $R$ with $Y$ held fixed. When doing so, in the same way that a likelihood function does not give a probability
distribution, neither should one expect `$P(R|Y_{obs},Y_{mis})$,' when considered to be a function of $Y$ with
$R$ held fixed, to give a probability distribution of any kind.
When one likewise comes to interpret the function `$P(R|Y_{obs})$' on the right hand side of the standard
equation, however, a problem arises. When the missingness mechanism is not-MAR, it is not immediately clear how this
function should be defined, and it is difficult to locate such a definition anywhere in the literature. Defining this function
so that the notation `$P(R|Y_{obs},Y_{mis})=P(R|Y_{obs})$' states a mathematical condition equivalent to MAR as
defined by Rubin\;(1976) is taken up in Section~\ref{Sect:DefinitionForPRGivenYobs}. Irrespective of the
definition, we stress that just like `$P(R|Y_{obs},Y_{mis})$,' it is important to understand that `$P(R|Y_{obs})$' does
\textbf{not} denote a conditional probability distribution for~$R$. Seaman\;et.\,al.\,(2013,\,p.\,260) point out a
difficulty that arises if one tries to interpret `$P(R|\,Y_{obs})$' in this way.

\section{A definition for $P(R|\,Y_{obs})$}
   \label{Sect:DefinitionForPRGivenYobs}
For the purpose of this section, we suppose that $P(R\,|\,Y)$ denotes a missingness mechanism
$g(\mathbf{r}\,|\,\mathbf{y})$ (see Section~\ref{Sect:Densities}), and we wish to determine whether or not
$P(R\,|\,Y)$ is MAR with respect to some realised values~$(\mathbf{y},\mathbf{r})\in\mathcal{Y}\times\mathcal{R}$.
To do so, we restrict the domain of $P(R\,|\,Y)$ to the observed data event
$\Omega_{(\mathbf{y},\mathbf{r})}\subset\mathcal{Y}\times\mathcal{R}$
and consider the range of $P(R\,|\,Y)$ when restricted to this section of its domain:
\begin{equation}
  \mathcal{S}\;=\;
      \{\,
      g(\mathbf{r}\,|\,\mathbf{y}_*):\, (\mathbf{y}_*, \mathbf{r}) \in
      \Omega_{(\mathbf{y},\mathbf{r})}
      \} \;\subset\;
      \mathcal{Y}\times\mathcal{R}.
   \label{Eq:RangeOfRGivenYObsYmis} 
\end{equation}
That is, we consider the set of values $\mathcal{S}$ that $P(R|Y_{obs},Y_{mis})$ takes on when setting
$R=\mathbf{r}$ and $Y_{obs}=\mathbf{y}^{ob(\mathbf{r})}$ and letting $Y_{mis}$ vary over all possible values.
We reiterate that in (\ref{Eq:RangeOfRGivenYObsYmis}) we are not interpreting $g(\mathbf{r}\,|\,\mathbf{y}_*)$ to
be a conditional density for $\mathbf{r}$ given $\mathbf{y}_*$. Rather, we interpret $g$ to be a function of the two
vector variables $\mathbf{r}$ and $\mathbf{y}_*$ (defined on all of $\mathcal{Y}\times\mathcal{R}$) which we
restrict to the subset $\Omega_{(\mathbf{y},\mathbf{r})}\subset\mathcal{Y}\times\mathcal{R}$ of its domain.
 
By definition $P(R\,|\,Y)$ is MAR with respect to $(\mathbf{y}, \mathbf{r})$ if, and only if, the set
$\mathcal{S}$ in (\ref{Eq:RangeOfRGivenYObsYmis}) contains only a single value. So, to distinguish between MAR
and non-MAR mechanisms, it suffices to define `$P(R\,|\,Y_{obs})$' to be
\begin{equation}
   P(R\,|\,Y_{obs}) \,:=\, \text{sup}\;\mathcal{S}.
   \label{Eq:RGivenYObs} 
\end{equation}
The supremum is well defined because $\mathcal{S}\subseteq [0, 1]$ is a bounded set. When $P(R\,|\,Y)$ is MAR,
one has $P(R\,|\,Y_{obs},Y_{mis})=P(R\,|\,Y_{obs})$ for all $Y_{mis}$ values because both sides equal the same
constant value, and when $P(R\,|\,Y)$ is not MAR, there will be at least one $Y_{mis}$ value for which
$P(R\,|\,Y_{obs},Y_{mis})\neq P(R\,|\,Y_{obs})$ because the right hand side is a constant (has the same value for
all~$Y_{mis}$ values) while the left hand side is not.

The definition for $P(R\,|\,Y_{obs})$ then extends naturally to all of $\mathcal{Y}\times\mathcal{R}$ by repeating
(\ref{Eq:RangeOfRGivenYObsYmis}) and~(\ref{Eq:RGivenYObs}) for all possible observable data events (see
Section~\ref{Sect:ODEs} for more details).

\vspace*{2mm}
\noindent
\textbf{{Note}.}
It makes no difference if we take the infimum instead of the supremum in~(\ref{Eq:RGivenYObs}).
More generally, we could take various combinations of supremum and infimum for the different missingness patterns,
so there are many different ways that the notation $P(R|\,Y_{obs})$ could be defined. Specifically, there is more than
one way to define the right hand side of $P(R\,|\,Y_{obs},Y_{mis})=P(R\,|\,Y_{obs})$ so that it becomes a
mathematical statement equivalent to MAR as defined by Rubin\,(1976). 
\hfill $\qedsymbol$

\vspace*{2mm}
\noindent
\textbf{Note.}
The significance of the function $P(R\,|\,Y_{obs})$ when the missingness mechanism is everywhere MAR is that it is
well-defined on the set
$\Omega_{ob} = \{ (\mathbf{y}^{ob(\mathbf{r})}, \mathbf{r}) : (\mathbf{y}, \mathbf{r}) \in \Omega \}$ of
observable data, and it carries all of the information about the dependence of $R$ on~$Y$.
So, if one did have complete information about $f(\mathbf{y})$, then this together with the function
$P(R\,|\,Y_{obs})$ would be sufficient to fully reconstruct the entire full distribution for~$(Y, R)$. That is, the
correct general interpretation of MAR is that no information about the missingness mechanism is lost through missing
data when $P(R\,|\,Y)$ is everywhere MAR.
\hfill $\qedsymbol$

\vspace*{2mm}
\noindent
\textbf{Note.}
In the literature, one sometimes comes across statements that seem to suggest distributional information about
$f(\mathbf{y})$ is not lost when the missingness mechanism is everywhere~MAR. While this seems to be correct
when the only source of missingness is dropout in longitudinal studies~(Molenberghs\;et.\,al.\,1998), it would seem
not to be a valid conclusion for general patterns of missingness. However, it is not easy to locate in the literature the
correct implications. Possibly the most one could conclude in general is that an everywhere MAR missingness
mechanism frees the analyst from having to model~$R$ explicity, and both likelihood theory and multiple imputation
rely on the correct specification of the full model $\{f_\theta(\mathbf{y}):\theta\in\Theta \}$ for the data vector~$Y$
for their validity. That is, these methods work by replacing any information about $Y$ that is lost through missing data
with the assumption that one of the~$f_{\theta_0}(\mathbf{y})$ from the model is the correct distribution for~$Y$.
\hfill $\qedsymbol$

\section{Discussion}
    \label{Sect:Discussion}
There has been some confusion in the literature regarding the correct definition(s) of MAR.
This is partly because different strength definitions are required for direct likelihood inference compared to frequentist
likelihood inference. This was clarified by Seaman\;et.\,al.\,(2013), with the weaker and stronger forms of the definition
being called realised and everywhere~MAR, respectively (see Section~\ref{Sect:MAR} for further details). Additionally,
MAR has been misinterpreted in the literature as a form of conditional independence (Carpenter and
Kenward,\;2013\;p.\,12, Fitzmaurice\;et.\,al.,\;2004\;p.\,381 and Molenberghs\;et.\,al.,\;2015\;p.\,8 for example). This
misinterpretation has been pointed out recently (Mealli and Rubin\;2015), where the authors refer to MAR as being
\textit{``incorrectly stated or inappositely redefined in the literature''}. This criticism seems unwarranted because the
authors provide no explanation as to why this misunderstanding has arisen in the first place.

As discussed in Section~\ref{Sect:Obstacles}, to interpret (5.13) as intended by the authors, the reader must be
aware that neither $f(R\,|\,Y_{obs})$ nor $f(R\,|\,Y_{obs}, Y_{mis})$ are being interpreted as probability
distributions. In the early textbooks (Little and Rubin\;1987; Schafer\;1997) this does not seem to have been pointed
out, although something to that effect is stated by Rubin (1987\;p.\,50). Additionally, unlike what is the case with the
likelihood function, neither Little and Rubin (1987) or Schafer (1997) changed the notation from a conditional
distribution to something that would make it clear that $f(R\,|\,Y_{obs}, Y_{mis})$ is not being interpreted as a
conditional probability density function. Thirdly, no definition for the right hand side of (5.13) seems to be given in
either edition of Little and Rubin (1987, 2002), or by Schafer\;(1997), and unless one knows how to define this, (5.13)
does not give well-defined condition for~MAR. We suggest that under these circumstances, confusion about the
correct interpretation of (5.13) seems quite reasonable.

We have remedied these gaps by explaining carefully why neither $f(R\,|\,Y_{obs}, Y_{mis})$ nor $f(R\,|\,Y_{obs})$
in (5.13) are to be interpreted as probability distributions , and we have shown how the function $f(R\,|\,Y_{obs})$
can be defined in an easy-to-understand manner so that the standard notation can be related back to Rubin\;(1976).
We hope that filling these gap will enable readers to approach the literature on statistical methods for incomplete data
with a little more confidence and greater understanding.

\section{References}
\begin{itemize}
   \item[Ash, R.B. and Dol\'{e}ans-Dade, C.A. (2000).]
      \emph{Probability and Measure Theory}.
      San Diego, California: Academic Press.

  \item[Carpenter, J.R. and M.G Kenward, M.G. (2013).]
    \emph{Multiple Imputation and its Application}.
    Chichester, West Sussex: Wiley.

  \item[Fitzmaurice, R.M., Laird, N.M. and Were, J.H. (2004).]
    \emph{Applied Longitudinal Analysis}.
    Hobokan, New Jersey: Wiley-Interscience.

   \item[Galati, J.C. (2019).]
      When is $Y_{obs}$ missing and $Y_{mis}$ observed?
      arXiv: 1811.04161\;{v3}.

   \item[Little, R.J.A and Rubin, D.B. (1987).]
      \emph{Statistical Analysis with Missing Data}.
      1st ed. New York: Wiley.

   \item[Little, R.J.A and Rubin, D.B. (2002).]
      \emph{Statistical Analysis with Missing Data}.
      2nd ed. Hobokan, New Jersey: Wiley-Interscience.

   \item[Mealli, F. and Rubin, D.B. (2015).]
     Clarifying missing at random and related definitions, and implications when coupled with exchangeability,
     \emph{Biometrika}, \textbf{102}:4, 995--1000,
     https://doi.org/10.1093/biomet/asv035;
     \textit{Correction:}
     \emph{Biometrika}, (2016) \textbf{103}:2, 491,
     https://doi.org/10.1093/biomet/asw017.

   \item[Molenberghs, G., Michials, B., Kenward, M.G., Diggle,P.J. (1998).]
    Monotone missing data and pattern-mixture models,
    \emph{Statistica Neerlandica}, \textbf{52}, 153--161.

   \item[Molenberghs, G., Fitzmaurice, G., Kenward, M.G., Tsiatis, A., Verbeke, G. (eds.)] (2015).
      \emph{Handbook of Missing Data Methodology}.
      Boca Raton FL: Chapman \& Hall/CRC.

   \item[Rubin, D.B. (1976).]
      Inference and missing data.
      \emph{Biometrika}, \textbf{63}:3, 581--592.

      https://doi.org/10.1093/biomet/63.3.581.

   \item[Rubin, D.B. (1987).]
      \emph{Multiple Imputation for Nonresponse in Surveys}.
      New York: Wiley, 1987.

   \item[Schafer, J.L. (1997).]
      \emph{Analysis of Incomplete Multivariate Data},
      Boca Raton FL: Chapman \& Hall/CRC.

   \item[Seaman, S., Galati, J., Jackson, D., Carlin, J. (2013).]
      What is Meant by `Missing at Random'?,
      \emph{Statist. Sci.}, \textbf{28}:2, 257--268,
      doi:10.1214/13-STS415.

   \item[ Shorack, G.R. (2000).]
      \emph{Probability for Statisticians},
      2nd ed.
      New York NY: Springer-Verlag.
\end{itemize}

\end{document}